# Sucrose ester surfactants: current understanding and emerging perspectives

Diana Cholakova, Slavka Tcholakova*

*Department of Chemical and Pharmaceutical Engineering*

*Faculty of Chemistry and Pharmacy, Sofia University,*

*1 James Bourchier Avenue, 1164 Sofia, Bulgaria*

\*Corresponding author:

Prof. Slavka Tcholakova

Department of Chemical and Pharmaceutical Engineering, Sofia University

1 James Bourchier Ave., Sofia 1164

Bulgaria

E-mail: sc@lcpe.uni-sofia.bg

Tel: +359 2 8161639

Fax: +359 2 9625643

**Abstract**

Sucrose esters (SEs), derived from sucrose and fatty acids, are biodegradable and non-toxic surfactants increasingly favored as substitutes for petrochemically-synthesized ones in food, cosmetics, and pharmaceuticals. SEs provide versatile hydrophilic-lipophilic properties, determined by the degree of sucrose esterification ranging from one to eight. The length of the fatty acid residues further influences the phase behavior of SEs, allowing creation of tailored formulations for specific applications. This review provides insights about our current understanding of the SEs phase behavior, their aggregation in aqueous and oily solutions, and its correlation with formulation outcomes. Furthermore, an overview of recent studies investigating SEs in various colloidal systems, incl. emulsions, foams, oleogels, and others, is provided. Novel concepts are discussed alongside future research directions, emphasizing the SEs potential as sustainable, functional ingredients.

# 1. **Introduction**

Sucrose fatty acid esters, commonly known as sucrose esters (SEs), are nonionic surfactants derived from the esterification of sucrose with fatty acids (FAs) [1-3]. Their synthesis is relatively simple, utilizing renewable and inexpensive natural resources [2]. These surfactants are considered as biodegradable and bio-compatible, they are tasteless [2] (particularly when the hydrophobic tail contains ≥ 12 C-atoms [4]), odourless, and non-irritant for skin [3]. They exhibit antimicrobial activity (especially against gram-positive bacteria [4,6]), antitumor and antioxidant properties, anti-inflammatory effects, insecticidal properties, and may serve as drug permeability enhancers [1-12]. All these characteristics make SEs widely employed in cosmetics, food, and pharmaceuticals [1-3,10-12].

Despite their widespread use in many formulations over the years, the phase behavior of SEs in bulk and in solutions remains incompletely understood, and research aimed to determine their foaming and emulsifying properties continues [13-18]. The major aim of this review is to summarize the current understanding of the main unique properties of SE surfactants with respect to their phase behavior in solutions, their adsorption, foaming and emulsion properties, and when possible to compare them to those of the petrochemical nonionic surfactants widely used in practice, e.g. ethoxylated alcohol ethers (Brijs) and ethoxylated sorbitan esters (Tweens).

Commercially available SEs typically appear as mixtures of mono-, di- and poly-esters with varying purities and hydrophobic chain lengths. These esters are commonly denoted with a single letter followed by a three- or four-digit number, such as S570 or L1695. The letter represents the hydrophobic chain length: C for capric ($C_{10}$), L for lauric ($C_{12}$), M for myristic ($C_{14}$), P for palmitic ($C_{16}$), and S for stearic ($C_{18}$). The last two digits indicate the purity of the tails, for example S570 contains about 70% stearic tails, whereas the other 30% are shorter or longer. The first digit in three-digit numbers or the first two digits in four-digit numbers represent the hydrophilic-lipophilic balance (HLB) of the sucrose ester surfactant, which is determined by the content of various molecular species in the mixture. This notation is also adopted in the present article.

The review is structured as follows: First, the phase behavior of sucrose esters powders is described, followed by an examination of the phase behavior of aqueous solutions of sucrose esters. The subsequent sections present the recent development in the utilization of SEs, including their foaming properties, emulsifying ability, preparation of foamed emulsions, oleogels and oleofoams. Finally, the review concludes with a summary of the main findings and identifies key scientific questions that remain open for future research.

## 2. Sucrose esters in solid state

Sucrose molecule possesses eight hydroxyl groups, allowing for a range of esterification degrees from one (monoesters) to eight, see Figure 1a,b. The fatty acid residues in mono-, di-, and tri-esters preferentially attach to the oxygen atoms positioned at 1', 6' or 6 sites, as these hydroxyl groups react preferentially [7,19,20]. Nonetheless, the formation of higher esters is also possible depending on the sucrose to FA ratio used during the synthesis. This variability, coupled with the diverse chain lengths of FAs that can be attached to the sucrose molecule, yields a mixture of molecules with a wide range of hydrophilic-lipophilic balances (HLB), making SEs suitable for both water- and oil-continuous formulations [1-3,7]. The increase of the concentration of di- and higher esters leads to the formation of more hydrophobic emulsifiers, whereas higher HLB values are observed when the monoester content predominates. We note that the HLB of SEs given by the producers is typically not calculated according to the classic definition of Griffin (HLB = $20 \times M_{hydrophilic\ part}/M_{total}$), but rather as: HLB $\approx 20\times$[monoesters content, %]/100 [21]. An approximately equal sign has been used because this definition usually gives about 1-unit deviation between the HLB stated by the producers and the one calculated using the accurate monoesters content in the sample.

The phase studies of pure mono- and diesters showed three main stages upon heating [7,22*,23], see Figure 1c,d: the SE molecules arranged in solid phases with lamellar structure at low temperatures (except for $C_8SE$ monoesters which arrange in disordered hexagonal columnar phase). Upon heating, these phases transformed to the less ordered smectic A* phase, i.e. liquid crystalline phase in which the molecules form layered structure, but no ordering exist within the layers, see the inset in Figure 1c. The asterisk symbol denotes that the material is chiral [7]. Finally, at even higher temperatures isotropic liquid phase was formed, although for the diesters partial preservation of the inner molecular ordering was reported [23].

Solid-to-solid phase transitions and formation of smectic A* phase was observed for all studied pure sucrose monoester homologues (1', 6 and 6'-substituted systems with $n = 12$, 16 or 18) [7]. Interestingly, the reported data show that the effect of position at which the FA chain is attached to the sucrose moiety is even bigger than the one observed from the increase of the FA chain length. In particular, the phase transition temperature increased from 46.2°C to 65.8°C when changing the position of the FA attachment from 1' to 6 for SEs with lauroyl chain ($n = 12$, $C_{12}SE$). In contrast, this phase transition temperature increased only by 6.6°C for 6-SE isomers (from 65.8°C to 72.4°C) and 7.6°C for 1'-SE isomers (from 46.2°C to 53.8°C) when increasing the FA chain length from 12 to 16 C-atoms [7]. The solid-to-smectic A* phase transition temperature decreased upon further increase of the FA length to 18 C-atoms and became 48.5°C for 1'-SEs and 39.3°C for 6-SEs [7], see

Figure 1c. This unexpected result was not investigated in more details, but it may be due to some impurities present in the longer SEs.

The smectic A* phase remained stable in a wide temperature interval for $C_{12}SE$, $C_{16}SE$ and $C_{18}SE$. The clearing point (temperature at which the smectic A*-to-liquid phase transition takes place) depended strongly on the SEs chain length increasing from ca. 180°C for $n = 12$ to 210-225°C for SEs with $n = 16$ or 18 C-atoms, see the filled symbols in Figure 1c [7]. The position at which the FA moiety was attached to the sucrose molecule affected the melting temperature, see Figure 1c. In particular, $T = 179.5°C$ was reported for the melting of mono-6-O-dodecyl sucrose, whereas the temperatures at which the isotropic liquid phase was detected in mono-6'-O-dodecyl sucrose was 204.5°C, respectively [7].

The shortest SEs investigated, with acyl chain $n = 8$, were found to exhibit disordered hexagonal columnar phase at room temperature, whereas the isotropic liquid phase was formed upon heating to 90-105°C without formation of smectic A* phase [7].

The diesters phase behavior in bulk state has been characterized in Refs [22*,23] for specially synthesized diesters with high purity. Similarly to the monoesters, the phase transition temperatures for sucrose diesters were found to depend strongly on the position at which FA are attached to the sucrose. Solid-to-smectic A* phase transition temperatures for 6,6'-diesters were reported to be higher than those for the 1',6'-diesters, compare the pink and dark blue points of same type in Figure 1d [22*]. Furthermore, the investigation of mixed chain diesters (with one lauric and one palmitic chain, thus average C-number = 14, star symbols in Figure 1d) showed that the phase transition temperatures for these diesters were even slightly lower compared to the phase transition temperatures of the pure dilauroyl esters [22*]. This result highlights the destabilizing role of the chain mixing over the stability of ordered phases, as typically found for all different classes of organic molecules [24*].

The melting of the more disordered smectic phase for diesters proceeded at $T \approx 160\text{-}170°C$, Figure 1d [22*,23]. The enthalpy of this second phase transition was relatively small, $\Delta H \approx 1\text{-}2$ J/g [22*,23]. Interestingly the specific temperature at which this phase transition was observed did not depend significantly on the chain length, see Figure 1d, whereas significant dependence is reported for the monoesters, Figure 1c. However, only the higher order peaks in SAXS disappeared for diester samples completely after this transition, while the primary (001) reflection remained present even at $T = 200°C$ [22*]. This indicates a presence of substantial ordering of diester molecules even in the liquid state, most probably governed by the presence of H-bonds between the sucrose moieties. Similar ordering in liquid state is known to exist for triglyceride molecules, although the exact reason behind it remains debatable and several different explanations have been proposed in the literature [24*]. However, to the best of our knowledge, whether such order will also exist in the liquid state of sucrose monoesters and what is the reason behind it for the sucrose diesters, have not been investigated so far.

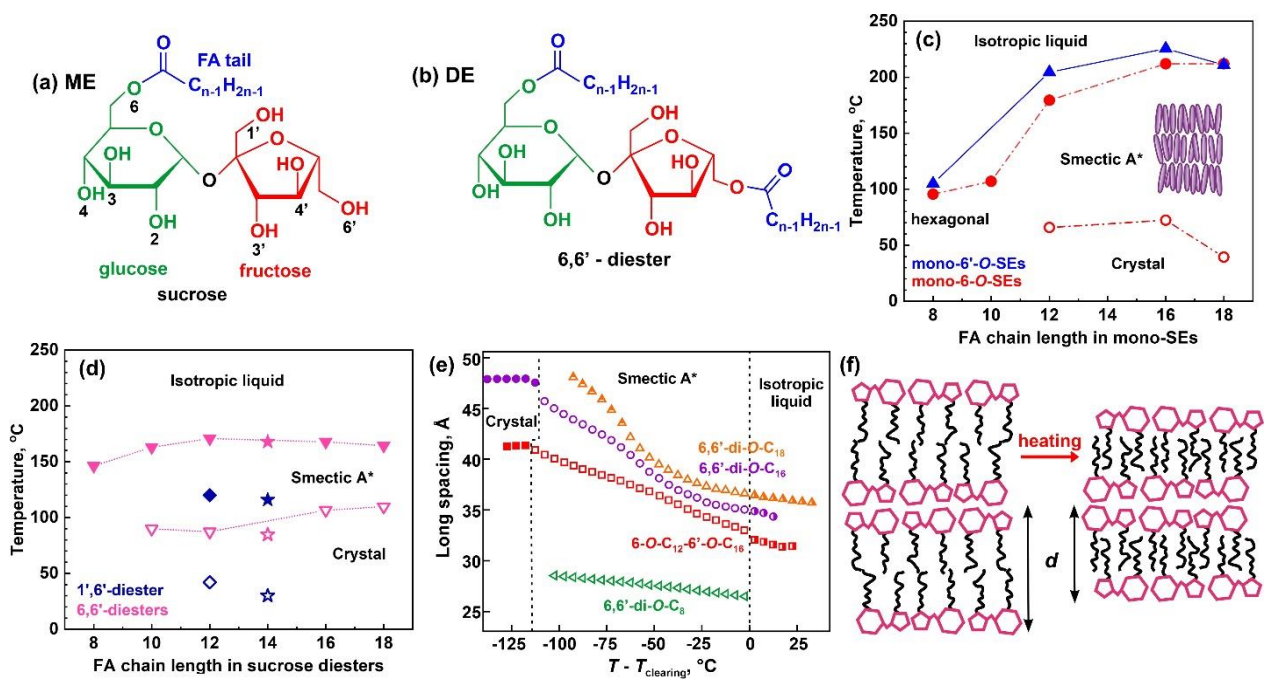

**Figure 1.** **(a,b)** General structure of sucrose: **(a)** monoester (ME) and **(b)** diester (DE) molecules. The numbers placed next to part of the oxygen atoms denote the positions of the hydroxyl groups within the sucrose molecule. **(c,d)** Phase diagrams for sucrose monoesters **(c)** and diesters **(d)** of high purity. The empty symbols represent the crystal-to-smectic A* phase transition temperatures, whereas the filled symbols show the smectic A*-to-isotropic liquid phase transitions. The color legend is explained on the figures. Note that the diesters shown in (d) for *n* = 14 are not pure diesters with myristic chains, but diesters with one lauric and one palmitic chain (*n* average = 14). Data taken from Refs [7,22*]. **(e,f)** The long spacing in sucrose diesters in smectic phase is found to decrease upon temperature increase. It is explained with a bilayer to monolayer transition upon heating. $T_{clearing}$ denotes the temperature at which the smectic A*-to-liquid phase transition takes place. Adapted from Ref. [22*].

Along with the transition temperature the information about the self-organizing structures of pure sucrose monoesters [7] and pure diesters [22*,23] was provided. In particular, lamellar structural organization with a long spacing of 4.13 nm was determined at $T$ = 30°C for mono-6-O-octadecanoylsucrose. Furthermore, the authors detected two WAXS peaks showing characteristic distances ≈ 4.3 Å and 3.8 Å, which were attributed to the scattering of aliphatic chains and the array or organized head groups, respectively [7]. The analysis of the inner molecular structure of the sucrose diesters revealed interesting results for the long spacing dependence from temperature, see Figure 1e,f [22*,23]. The long spacing for the smectic A* phase was found to decrease upon temperature increase. This behavior is opposite to the one typically found for other classes of molecules, including for the rotator phases of linear alkanes which are somewhat similar to the smectic liquid crystalline phases [24*-26]. This was attributed to a bilayer to monolayer polymorphic phase transition proceeding upon heating while the material is organized into the liquid crystalline smectic A* phase [22*,23]. It was proposed that this transition involves the formation of folded interdigitated structure at higher temperatures from the initially non-interdigitated one observed at lower temperatures [22*].

The phase behavior of several commercial SEs with FA tails containing 16 ("P", palmitic), 18 ("S", stearic) or 22 C-atoms ("B", behenic) and different monoester content, varied between ca. 15% for B370 SE (HLB ≈ 3) and 80% for P1670 SE (HLB ≈ 16) was investigated in Ref. [27], see Table 1. The authors showed that SE with intermediate and higher monoester (ME) content (≥ 50 %), exhibited glass transition upon the initial heating, while a melting process was observed for SEs with lower ME content (< 20 %), see Table 1. The glass transition and melting temperatures varied between ca. 50°C for palmitic SE and 77°C for the behenic SE. All studied samples exhibited more than one pronounced peak maximum [27]. However, it remains unclear whether these multiple maxima are related to the consecutive melting of the molecules with different degree of substitution, to the melting of molecules containing different FA residues or to the melting of SEs in which the FA residues are attached to different positions as shown in Ref. [7]. It was also shown that after the melting and solidification, SEs have partially amorphous structures which slowly crystallize in time. The crystallization was faster for SEs with lower ME content [27]. Furthermore, our recent DSC data obtained with similar SEs did not confirm that the SEs with higher monoester content exhibit glass transition, as crystallization peak with enthalpy similar to the one measured upon heating was observed upon subsequent cooling, see Supplementary Figure S6a in Ref. [28]. Therefore, further investigations will be needed to determine the reason for this discrepancy.

**Table 1.** Temperature range and enthalpy ($\Delta H$) of melting, and long spacings ($d$) for selected commercial sucrose esters. All data are taken from Ref. [27].

| SE | HLB | Monoesters, % | Melting range, °C | $\Delta H$, J/g | $d$, nm |
|---|---|---|---|---|---|
| **P1670** | 16 | 80 | 41-62 | 66.6 | 4.14 |
| **S1670** | 16 | 77 | 45-62 | 66.7 | 4.14 |
| **S970** | 9 | 48 | 36-65 | 75.7 | 5.78; 4.45; 2.00 |
| **S370** | 3 | 19 | 45-66 | 55.7 | 6.08; 2.00 |
| **B370** | 3 | 15 | 51-79 | 68.4 | 6.79; 4.82; 2.37 |

*"L" = lauric; "P" = palmitic; "S" = stearic; "O" = oleic; "B" = behenic; Number codes: the last two digits represent the purity of the FA chains, i.e. in this case all studied SEs contain ca. 70% of the respective FA; and the first (in the three-digits codes) or the first two digits (for four-digits codes) represent the HLB value given by the producer. The melting range data are determined upon the initial heating of the samples. SAXS data show the long spacings of the main peaks.*

The inner molecular arrangement determined from SAXS/WAXS measurements showed well pronounced peaks for the studied SEs samples, see data in Table 1. For the hydrophilic P1670 and S1670 only one main peak was observed corresponding to interlamellar distance $d \approx 4.14$ nm [27]. Note that the absence of any measurable difference in the long spacing shown by SE containing predominantly $C_{16}$ FA residues (P1670) and the one with $C_{18}$ chains (S1670) should be governed by the mixing behavior of the different chains and the presence of *gauche* bonds in the longer $C_{18}$ tails [24*]. This spacing is significantly longer than the length of $C_{18}$ tail which is $\approx 2.3$ nm (data for the chain length of linear *n*-octadecane [20]) and slightly shorter than the long spacing determined for stearic acid bilayer structure ($d \approx 4.39$ nm for B-form and 4.65 nm for the more disordered A-form) [29]. This is probably due to the presence of some fraction of shorter FA residues in the molecules and/or due to the formation of interdigitated bilayers as proposed in Ref. [22*].

Our investigation of palmitoyl-stearoyl SE (ca. 80% $C_{16}$ FA and 20% $C_{18}$ FA, with ca. 75% MEs) [28] showed lamellar structure with $d \approx 3.98$ nm and four well visible reflections at temperatures below the gel transition temperature for the investigated SE. At temperatures above ca. 43°C, the WAXS peak showing the aliphatic chains arrangement became much more scattered, indicating the melting of the surfactant tails (due to the formation of the smectic A* phase). As a result, the long spacing shifted to $d \approx 4.17$ nm, a value very similar to the one found in Refs. [7,22*]. The long spacing increased with the increase of the diester fraction in the sample, see Table 1. Furthermore, two or three well distinguished maxima were observed for the SEs with higher di- and polyesters content [27].

Finally, we will mention an interesting study characterizing the surface hydrophobicity of thin layers of sucrose ester surfactants with HLB $\approx 1$ (i.e. containing mainly di- and higher esters) and different FA chains [30]. The investigated layers were prepared by dissolving the studied SE in chloroform and then soaking a cover glass in it. Afterwards, the glass slides loaded with SEs were took out from the chloroform and it was allowed to evaporate in a fume hood. The three-phase contact angle measurements were performed with a water droplet placed over the studied glass slide and immersed in corn oil. The results showed two main trends [30]. The increase of the hydrophobic tail length was found to increase the hydrophobicity of the layer for both saturated and unsaturated SEs. In particular, the three-phase contact angles, $\theta$, measured through the water droplet, increased from $\theta \approx 120°$ for $C_{12}$SE to 130° for $C_{16}$SE, and 132° for $C_{18}$SE. Furthermore, the layers prepared with surfactants containing predominantly unsaturated FA residues (erucic acid, $C_{22:1}$, and oleic acid, $C_{18:1}$) were significantly more hydrophilic than layers prepared from saturated SEs, $\theta \approx 101°C$ for $C_{18:1}$SE and $\theta \approx 112°C$ for $C_{22:1}$SE [30]. These properties are important when considering which would be the best surfactant for a given application, as shown in sections describing foams and emulsions below.

## 3. Sucrose esters self-assembly in aqueous media

Sucrose ester molecules spontaneously self-assemble into supramolecular structures (micelles, vesicles, lamellar phases, etc.) once their concentration into the aqueous phase exceeds the critical micellar (aggregation) concentration (CMC/CAC). As expected for nonionic surfactants, the CMC values are found to decrease upon increase of the tail length and decrease with the decrease of percentage of MEs in the mixture [5,17**,31-33]. In particular, the CMC values reported in the literature for octyl acid derivatives vary in a wide range, between 2.3 and 30 mM [31-33], and for $C_{10}SEs$ values around 1.5-4 mM are reported [5,31,33]. We note that different experimental methods for CMC determination, as well as sucrose esters of different purity have been used in these experiments, which explains the scattering of the reported data. For the longer chain SEs, the available data is much more consistent and CMC values around 0.3 mM for $C_{12}SE$, 0.02 mM for $C_{14}SE$ and $C_{16}SE$, and ≤ 0.01 mM for $C_{18}SE$ are measured [5,31-33].

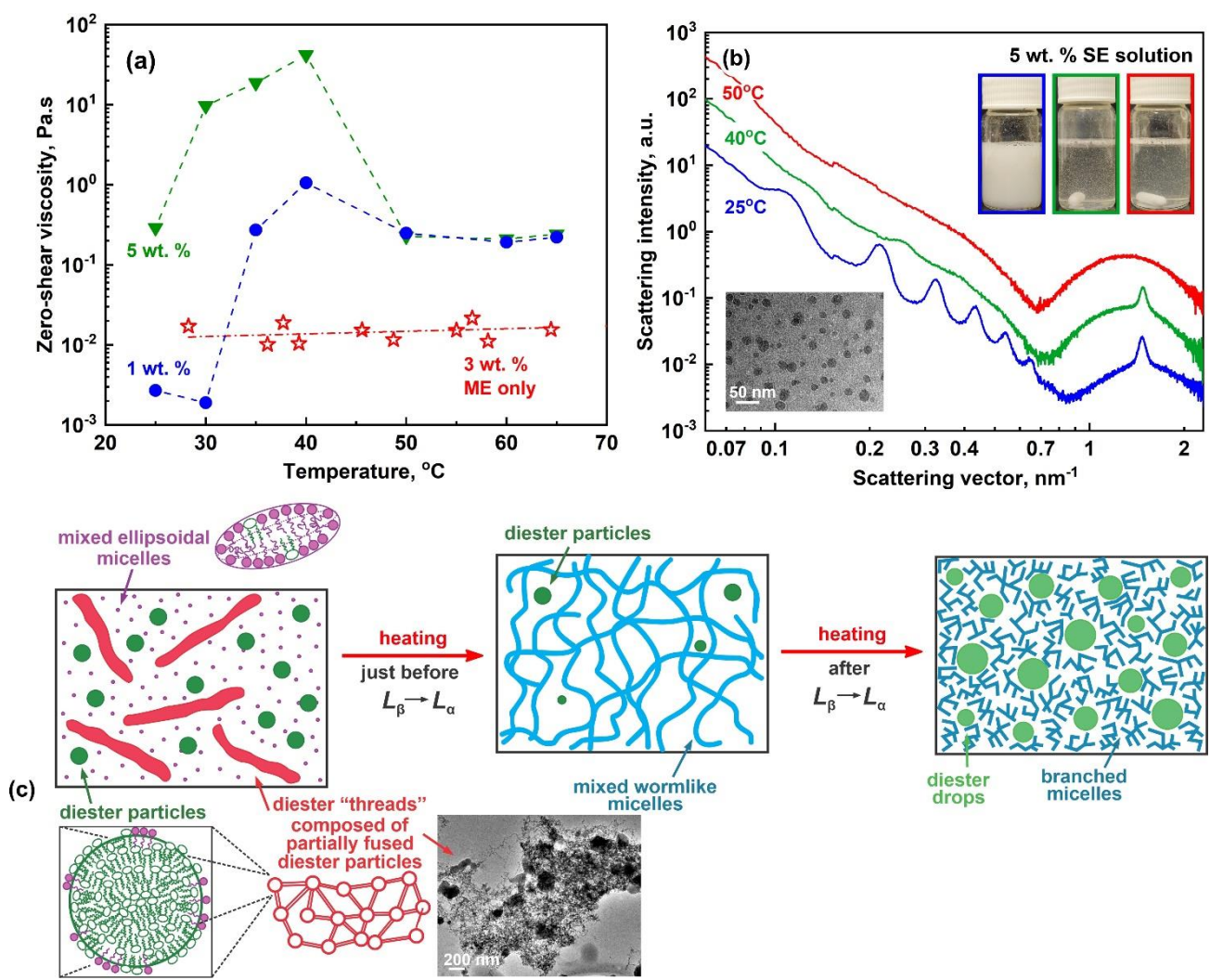

**Figure 2.** Rheological behavior of commercial SE surfactant ($C_{16}FA:C_{18}FA \approx 4:1$, ca. 75% MEs). **(a)** Rheology as function of temperature for 1 and 5 wt. % SE solutions (filled symbols). The empty stars represent the rheology of solution containing the monoester fraction separated after ultracentrifugation of 5 wt. % initial solution (ME concentration ≈ 3 wt. %). **(b)** SAXS/WAXS spectra

of 5 wt. % SE solution taken at different temperatures. Inset: picture of solution obtained at the same temperatures and cryo-TEM picture of diester particles causing the SAXS curve oscillations at 25°C. **(c)** Schematic representation showing the molecular rearrangements upon heating and cryo-TEM picture of the observed "threads". Adapted from Ref. [28]. See the main text for more details.

A linear dependence between the logarithm of CMC and the number of methylene units in the surfactant tail was established for pure 6-*O*-sucrose monoesters with chain length between 8 and 16 C-atoms in their tail [31]. This result is very similar to those reported usually for conventional nonionic alcohol ethoxylate surfactants, such as Brij or Tween [34]. The spherical or short rods micelles are formed at concentrations around CMC for SEs with $C_8$ to $C_{12}$ chain lengths [17**,31,33]. A linear decrease in the diffusion coefficients and a linear increase of the hydrodynamic radius for micelles of 6-*O*-monoesters is determined with the increase of methylene units for SEs with 10 to 18 C-atoms with $D \approx 7.37 \times 10^{-11}$ m$^2$/s for $C_{10}$SE and $D \approx 5.10 \times 10^{-11}$ m$^2$/s for $C_{18}$SE [35]. A slight decrease in the CMC values is reported upon temperature increase [5,17**].

The aqueous solutions of SEs exhibit lyotropic properties, depending on both surfactant concentration and temperature [19]. Intriguing rheology behavior has been reported for relatively diluted (≈ 1-10 wt. %) SEs solutions – the viscosity increases upon increase of the temperature up to a given value, and then it starts to decrease [28,36-38]. Note that for the typical nonionic surfactants, usually the viscosity decreases monotonically upon heating [34]. The opposite behavior has been observed in few very special surfactants/surfactant combinations and it is related to the continuous dehydration of the ethoxy groups (-O-CH$_2$-CH$_2$-) comprising the hydrophilic head of the nonionic surfactants [39-41]. For the SEs, however, the mechanism determining the observed viscosity increase was found to be different [28].

In a recent study, we investigated the phase behavior of SE surfactant comprising ca. 80% monoesters and 20% diesters, with $C_{16}$:$C_{18}$ chains ≈ 4:1 [28]. Illustrative results, showing the rheological behavior of the aqueous solutions of this surfactant, are shown in Figure 2a. The analysis of this non-trivial phase behavior revealed that it is mostly governed by the presence of diesters molecules. In contrast, when they were removed from the solution, it behaved as a typical diluted solution of nonionic surfactant for which the viscosity does not change significantly upon temperature increase, see the empty star symbols in Figure 2a [28].

The role of the diester molecules was revealed to be as follows, see schematics in Figure 2c [28]: at low temperature, the alkyl chains of the molecules are solid-like ($L_β$ phase) and they cannot arrange in joint structures with the monoester molecules. Hence, they phase separate as nanometer-sized particles with $d ≈ 60$ nm which exist as individual particles or build network of partially fused diester particles, making the solution very turbid at 25°C, see Figure 2b,c. As the content of diester molecules is relatively low, the solution has low viscosity. These particles/threads co-exist with slightly elongated micelles, containing predominantly the monoester molecules. Upon heating, however, the packing of the alkyl chains become disrupted due to the increasing number of defects related to the increased energy of the system, thus a rearrangement is observed. The main part of the diester molecules become included in mixed wormlike micelles together with monoester molecules, Figure 2c. This leads to increase in the viscosity of the solution by several orders of magnitude and formation of gel-like transparent systems at very low surfactant concentrations (above ca. 1-2 wt. %). Zero-shear viscosity of ca. 2 Pa.s is observed for 2 wt. % SE solution, whereas ca. 40 Pa.s viscosity is obtained for 5 wt. % solution at $T = 40°C$. Further increase of temperature leads to fluidization of the tails. As a result, part of the diesters escape from the mixed micelles and phase separate as drops in the solution. The viscosity of the solutions decreases, but remains higher as compared to the values measured at 25°C. Furthermore, no significant dependence from the surfactant concentration is observed for the zero-shear viscosity at $50°C \leq T \leq 70°C$. This can be explained with the formation of mixed branched micelles which coexist with diester droplets [28].

In the context of these new results, the concept about cloud point of mixed SEs existing in the literature [42] could be slightly revised. The cloud point for nonionic surfactant is defined as the temperature above which the surfactant aggregates begin to precipitate due to the progressive dehydration of the hydrophilic heads. This leads to phase separation between the surfactant and water [43]. Generally, the cloud point increases with the increase of the size of hydrophilic head and with the decrease of the length of the hydrophobic tail of surfactant molecule [43,44]. As all SE possess the same hydrophilic headgroup, their cloud point will be solely determined by the number and length of the hydrophobic chain(s).

The cloud points of alcohol ethoxylate surfactants have been widely studied and it is well established that they increase relatively monotonically with the increase of the HLB value, see for example Figure 1 in Ref. [42] and Ref. [44]. Interestingly, Ref. [42] reported a significantly different behavior for SEs. In particular, Figure 1 in Ref. [42] demonstrates that for $C_{12}$-tail mixed sucrose ester surfactants the minimal increase in the content of sucrose di- and tri-esters (resulting in the HLB decrease from ca. 11.4 to 11) leads to abrupt change in the temperature at which turbid solutions are obtained. Values higher than 100°C are reported for the more hydrophilic compositions (HLB $\gtrsim$ 11.4), whereas it decreases drastically to temperature below 20-25°C for the more hydrophobic ones (HLB = 11). This temperature is interpreted as being the "cloud point" of the studied surfactants.

Considering the role of the mono-to-diesters ratio for the overall phase behavior of the SEs' solutions [28], it becomes evident that the obtained results can be interpreted as follows. At relatively low di- and tri-esters content these higher esters are incorporated inside the monoesters' micelles. Upon increase of their concentration, however, which is expressed as a decrease of the HLB value, eventually the solubilization capacity of the monoesters is exceeded. This leads to precipitation of the excess amount of di- and triesters into individual nanoparticles/aggregates, which scatter the light and cause the turbid appearance of the solution. The monoesters and part of the higher esters most probably remain dissolved, however, as shown in Figure 2b for the studied $C_{16}SE$. Therefore, this precipitation temperature is not exactly the typical cloud point, but rather it is governed predominantly by the amount of higher esters and the solubilization capacity of the monoesters. For more hydrophilic SEs (HLB > 11.4) the "cloud point" reported is higher than 100°C [42], as expected for single-chain hydrophilic $C_{12}$-surfactants. Further studies with purified SEs are needed to determine more precisely their cloud points and to compare their behavior with that of the typical nonionic surfactants.

No significant differences were observed between the rheological behavior exhibited by a SE surfactant containing predominantly stearoyl [36] or palmitoyl [28] chains, except for the position of the peak maximum. While the maximal viscosity was observed at ca. 40°C for the palmitoyl SE [28], it shifted to ca. 47°C for the stearoyl SE [36]. This is related to the different phase transitions temperatures for these surfactants in aqueous solution. In particular, the phase transition for 2 wt % S1670 solution measured by DSC is observed at 47°C, while for P1670 it is observed at 40°C [16*]. The comparison between the gelling properties of S970 and P1670 made in Ref. [38] showed that under equivalent conditions, the gels prepared with S970 were with better visco-elastic properties compared to those prepared with P1670. Note that this result can be attributed to two different factors. From one side, both surfactants were investigated at the same temperature (37°C), while their melting points are different. From the other, this can be related to the effect of different monoester content in the studied SEs and different ratio between $C_{18}SEs$ and $C_{16}SEs$. Further studies are needed to resolve which is the leading factor for the obtained result and also what is the effect of the monoester content on the viscoelastic properties of the prepared gels.

Besides as a result from temperature variations in mixtures of sucrose monoesters and diesters, the possibility for formation of wormlike micelles with SEs has been also described in presence of various co-surfactants, incl. fatty alcohols [45], fatty acids [46], relatively hydrophobic nonionic surfactants (alcohol ethoxylates with small number of ethoxy units and monoglycerides) [47,48], and ionic surfactants [46,47]. In this case, the zero-shear viscosity of the prepared solutions exhibited a maximum as a function of the co-surfactant concentration, a behavior typically found for systems containing wormlike micelles. Zero-shear viscosities of 200-500 Pa.s were reported at 10 wt. % total surfactant concentration ($C_{16}SE$ + co-surfactant). This was explained with the preferential one-dimensional growth of the micelles in presence of additives due to the altered preferential packing of the molecules [45-48].

The investigations about the effect of FA chain in alcohols and acids showed that the increase of acyl chain led to increase in the maximal zero shear viscosity which can be observed, as far as the system remained in a liquid state, i.e. for $C_3$- to $C_9$-primary alcohols and $C_6$- to $C_{12}$-fatty acids [45,46].

Furthermore, the position of the zero-shear viscosity maximum shifted to lower co-surfactant concentrations when the co-surfactant chain become longer [45,46]. Interestingly, the addition of oily molecules to systems in which wormlike micelles were already formed had an alternative effect over the viscosity of the systems depending on the molecular structure of the oily molecules [48]. In particular, it was found that the small slightly polar aromatic hydrocarbon molecule m-xylene solubilize in the palisade layer of the wormlike micelles, thus extending the length of the wormlike micelles. This led to further increase in the viscosity of the solution. In contrast, the hydrophobic bulkier decane molecules solubilized in the core of the micelles, thus causing shortening of the wormlike micelles length and significant drop in the viscosity [48].

Experiments with higher SEs concentrations have been also performed and some phase diagrams are available in the literature, see e.g. Refs [19,21,49]. Formation of hexagonal and lamellar liquid crystals is usually reported at intermediate and high SE concentrations, respectively. Further detailed information can be obtained from the original papers.

4. **Foam properties of sucrose esters**

The foaming and interfacial properties of sucrose monolaurate (L1695) were studied by Husband et al. [50]. In this study the authors showed that this commercial surfactant (containing both laurate mono- and di-esters) exhibits better foamability than the individual pure sucrose monolaurate and sucrose dilaurate. Furthermore, to obtain similar foamability as the one observed with L1695, the authors needed to mix the pure monolaurate and dilaurate in molar ratio = 4:1. Although the authors did not observe any significant change in the surface and film properties with this mixture compared to the properties exhibited by the individual monolaurate, the foamability was significantly improved. Therefore, the synergistic effect between the mono- and diesters was attributed to the appearance of interactions between the molecules inside the solution and a reduction in the critical micellar concentration at this particular ratio [50].

The effect of diesters over the foams stability was further investigated in a recent study performed by our group [17**]. We showed that the diesters have stabilizing effect, which is especially pronounced at high temperatures. In particular, the foams stabilized by L1695 remained stable for ca. 10 min at 60°C, whereas the foams prepared with the widely used nonionic water-soluble alcohol ethoxylate surfactant (Brij L23) diminished for less than a minute under equivalent conditions. This was explained with the formation of dense adsorption layer from the diesters on the bubble surface, see Figure 3 [17**].

The length of the SE molecules and the foaming method used both influence the foamability and stability of the generated foams. For example, in a study using the sparger method (a slow foaming method) and SEs with chains between $C_{12}$ and $C_{18}$, it was found that increasing the SE tail length from 12 to 18 C-atoms resulted in an increase in bubble size in the produced foams from ca. 0.4 mm to 3.5 mm, respectively [51]. Additionally, foam life decreased when the surfactant chain length was increased, from 700 min for $C_{12}SE$ to 61 min for $C_{18}SE$ [51].

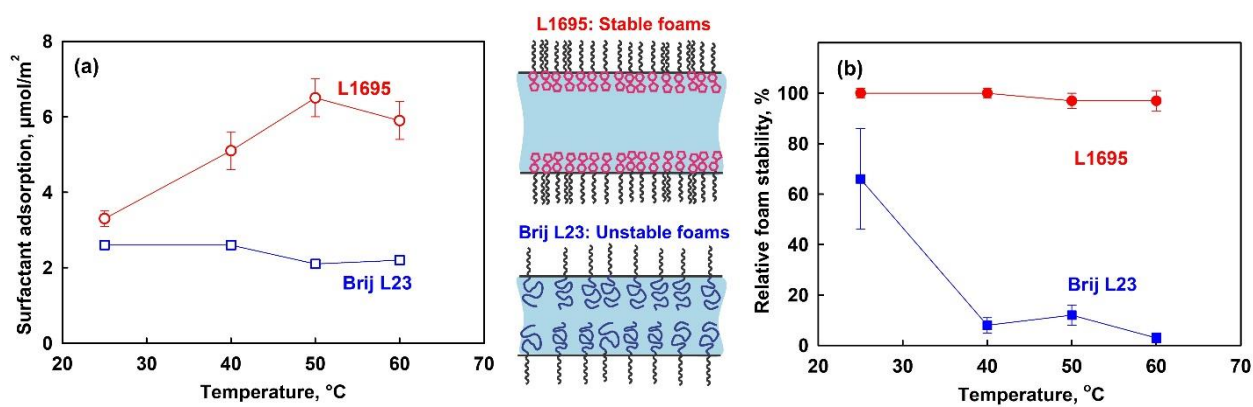

**Figure 3.** (a) Surfactant adsorption on air-water interface as a function of temperature. (b) Relative stability of foams, defined as percentage of remaining foam volume 10 min after the foam generation has been stopped. Foams are generated by Bartsch test from 1 wt. % surfactant solution at different temperatures. Red points: sucrose laurate (L1695); blue points: Brij L23 (polyoxyethylene (23) lauroyl ether). Data are adapted from Ref. [17**].

However, in a recent study we showed that using a planetary whipping mixer (Kenwood) for foaming, one can successfully produce foams with very small bubble sizes (5 μm) even when using long chain SE surfactants, such as P1670 or S1670 [16*]. These foams were prepared in presence of high sucrose concentrations and remained stable for more than one year at room temperature. The exceptional stability was explained with the formation of solid layer on the bubble surfaces which can change the surface tension of shrinking bubbles down to zero due to formation of wrinkles, thus decreasing the driving force for Ostwald ripening significantly. Similar exceptional stability for foams created from a mixture of glucose syrup and sucrose stearate mono- and diesters mixture was reported by Dressaire et al. [52**]. These authors demonstrated that the formation of a regular surface pattern is a thermodynamic signature for formation of an elastic, condensed surfactant phase, which also correlated with the extended stability of the systems stored at 4°C [52**].

Currently, it remains unclear what is the precise influence of glucose and sucrose in the aqueous phase on the properties of adsorption layers formed from alkyl sucrose esters and whether such exceptional stability can be achieved at lower sucrose/glucose concentrations. Further studies are needed to elucidate the impact of carbohydrates on SE properties. Additionally, given the varied foam generation methods used in these studies, their effect should be also evaluated in details. Nonetheless, we expect that stable foams can be generated with long chain SEs even at lower sucrose concentrations, as demonstrated previously with sorbitan monostearate (Span 60) [53]. This research has shown that Span 60 can stabilize aqueous foams not only in the presence of sucrose but also when dispersed in the aqueous phase, provided the appropriate foaming method is used [53].

## 5. Emulsions

Sucrose esters have been widely studied as emulsifiers, especially in food industry [1,12,54,55]. Owing to their varying hydrophilic-lipophilic properties, they can be used to stabilize both oil-in-water (o/w) and water-in-oil (w/o) emulsion systems. In this section, we will highlight only the newest development in this field to show which are the current directions under investigation.

### 5.1. Oil-in-water emulsions

The utilization of SE with intermediate HLB value (HLB ≈ 7) for preparation of highly concentrated o/w emulsions was investigated in Ref. [13]. The authors successfully prepared stable canola oil-in-water emulsions containing oil weight fraction, $\varphi_o = 0.5$, when at least 2 wt. % SE was used, whereas at lower concentrations the emulsions undergo coalescence upon storage. However, the emulsions prepared under similar conditions with soybean or olive oil were not stable. This was attributed to different triglyceride composition of the oils, but the detailed mechanism remains unclear. Furthermore, the authors also prepared emulsions with higher oil weight fraction, $\varphi_o = 0.75$, which is above the close packing value, making the droplets polyhedral in shape. The prepared high internal phase emulsions (HIPEs) had viscoelastic properties, but were unstable in all cases against coalescence after 3 months storage. The highest stability was once again found for the canola oil containing samples. The authors attributed the ability of the SE with intermediate hydrophilicity to stabilize o/w emulsions to the synergistic role of the SE monoesters which adsorb on the drops surfaces and of the di- and poly-esters which were proposed to arrange in multilamellar vesicles, which constituted the walls around the droplets, thus preventing them from coalescence [13]. It was shown that the destabilization takes place if the emulsions are stored at minus 20 °C instead of +5 °C upon thawing suggesting that a partial coalescence between drops takes place when the oil undergoes the crystallization. Further investigations are needed to confirm the proposed mechanism, to refine the effect of TAG oil composition and to propose HIPE formulations which will be stable upon prolonged storage.

In the study of Fernandes and co-authors [14], the SE and polysaccharides interactions were investigated in relation to preparation of stable emulsions with sunflower oil (SFO). The authors showed that the addition of guar or xanthan gum led to formation of SE-polysaccharide intermolecular complex, most probably via hydrophobic interactions or H-bonds formation, which led to increase in the interfacial tension of the mixed system as compared to the one measured when SE was only present [14]. The authors successfully prepared SFO-in-water emulsions, which remained stable upon storage. However, flocculation was observed in the samples containing gums, which was explained with a depletion mechanism related to the presence of non-adsorbent polysaccharides in the bulk continuous phase [14]. Similar conclusions about the complexation between SE and xanthan gum and their synergistic effect for stabilization of emulsions are described in Ref. [56].

Interestingly, in the study of Guo et al. [57], o/w micrometer-sized emulsions were tested for their potential to act as lubricants. The results show that o/w emulsions containing 2 wt. % SE, 6 wt. % octadecanol and 10 wt. % poly-α-olefin were able to decrease the coefficient of friction by ca. 34% compared to the one measured for the same system in presence of pure oil. Such high decrease was observed only when the emulsion was prepared under milder homogenization conditions to preserve the lyotropic liquid crystals formed by the sucrose ester emulsifier upon cooling. In contrast, the coefficient of friction decreases only by ca. 10% when the investigated emulsion was homogenized by ultrasound, so that the liquid crystal arrangement become broken. The better protective performance and anti-wear properties of the sample containing liquid crystalline phase was attributed to the homogeneous lubricating film formed by this emulsion [57].

*5.2. Micro- and nano-emulsions*

Rao and McClements studied the ability of hydrophilic sucrose ester SP90 (containing ca. 90% monoesters and ca. 96% palmitic chains) to form micro- and nano-emulsions with lemon oil [58,59]. Note that the microemulsions are thermodynamically stable and form spontaneously, whereas nanoemulsions are kinetically stabilized, but both types of emulsions contain nanometer-sized droplets. The results showed that microemulsions can be prepared when the surfactant-to-oil ratio is higher than 1 when heating the system to temperatures above 75°C [58,59]. This heating ensured the needed surfactant arrangement changes, thus leading to good oil solubilization inside the surfactant micelles once the emulsion was cooled back to ambient temperature (i.e. by the phase inversion temperature method, PIT) [58]. In contrast, when spontaneous emulsification by mass transfer was studied for similar systems, nanoemulsions were formed instead of microemulsions [60]. A review about the spontaneous emulsification methods is available in Ref. [61]. The prepared lemon oil microemulsions and nanoemulsions were stable at 25°C, but the increase of temperature to 40°C led to their destabilization, see Figure 4 [58]. This destabilization occurs from the fluidization of the surfactant adsorption layer when the temperature exceeds its melting point, enabling the incorporation of diester molecules into monoester micelles [28]. This incorporation significantly reduces emulsion stability.

The prepared microemulsions were stable at pH = 5 or 6, and when electrolyte was added to the system (up to 200 mM NaCl) [58]. A significant electric charge was determined for the nanoemulsions prepared in Ref. [58]. It was shown that the drops are highly negatively charged at pH

= 8, whereas the electrical charge decreased upon pH decrease, or upon addition of higher surfactant concentrations [58]. The negative charge for the aggregates formed in the aqueous solutions of non-ionic SEs has been observed also by other researchers [14,15,62]. It was also invoked to explain the very thick films formed from aqueous solutions of L1695 in Ref. [17**]. The negative charge was attributed to the presence of free fatty acids which appear as impurity for the SE upon their synthesis, or due to hydrolysis of the SE molecules in Ref. [58]. Another formulated hypothesis in Refs [15,17**] is that the adsorption of hydroxyl anions on the aggregates surface can be the reason for this electric charge. Note that the adsorption of $OH^-$ anions on the oil interfaces was shown to be operative for other nonionic surfactants such as $C_{16}EO_8$ which does not contain contamination of fatty acids [63]. However, currently there is no unambiguous conclusion what is the exact mechanism responsible for the observed negative charge for SE solutions. Therefore, this topic surely deserves further investigation to reveal the underlying reason.

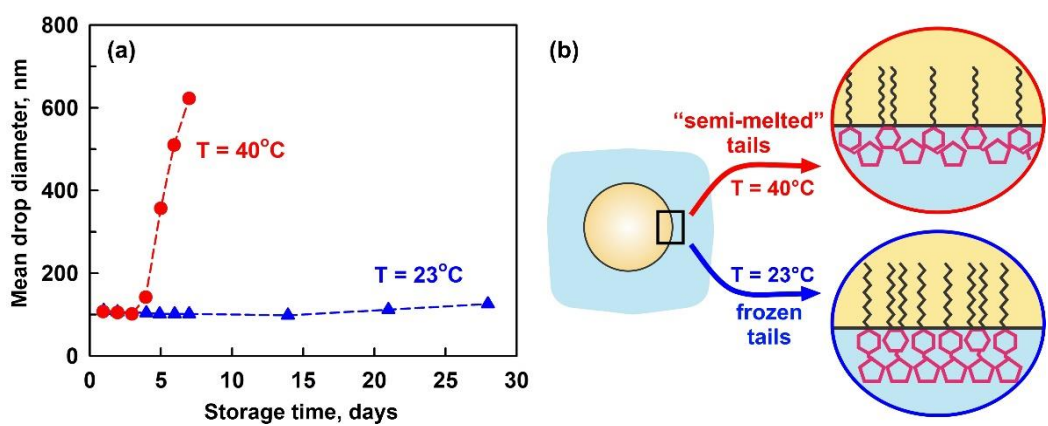

**Figure 4.** (a) Mean drop diameter evolution upon storage at two different temperatures 23°C (blue triangles) and 40°C (red circles) for nanoemulsions prepared with 1 wt. % sucrose palmitate (SP90), 10 wt. % lemon oil and 89 wt. % 10 mM sodium dihydrogenphosphate buffer pH = 7 (propylene glycol-to-water ratio = 1:2). Experimental data adapted from Ref. [58]. (b) Schematic representation of the oil-water interface, highlighting the partial melting of the SE tails at 40°C, which allows the diesters molecules exchange with the volume, thus decreasing significantly the stability of the observed emulsions and allowing the observed drop-drop coalescence.

Microemulsions which were stable at pH ≈ 3.5 were prepared in Ref. [64]. These o/w microemulsions were based on essential oils, isopropyl alcohol and SEs. They were designed so they can be potentially used for topical delivery of the drug fluconazole. However, further *in vitro* and *in vivo* studies will be needed to access the emulsions performance, bioavailability of the drug and the effectiveness of treatment of the skin fungal infections. Furthermore, the prepared microemulsions did not possess the needed viscosity for skin applications, hence a further viscosity increase will be needed [64].

The ability of three different SE to form nanoemulsions via the D-emulsification approach was studied in Ref. [65]. Under the investigated conditions (water content between 50 and 83 wt. %, olive oil and glycerol), the shortest chain lauric SE (L1695) was found to be able to form nanoemulsions with sizes in the range 230 – 370 nm in the largest compositional range. The longer $C_{18}$-chain SE surfactants with unsaturated (oleic) or saturated tails were found to be able to form nanoemulsions in narrower compositional range. The emulsions prepared with L1695 SE were found to be suitable for encapsulation of poorly water-soluble drug ibuprofen. They also enhance its absorption *in vivo* compared to systems in which the ibuprofen was encapsulated in microemulsion droplets or when it was directly dissolved into the oily phase [65].

### 5.3. Water-in-oil emulsions

The ability of hydrophobic sucrose esters to stabilize inverse, water-in-oil emulsions have been also demonstrated in several recent studies. Gao et al. [66], explored the ability of lauric and palmitic sucrose esters (L195 and P170) to stabilize w/o emulsions with water weight fraction $\varphi_w$ = 0.2, prepared with soybean oil (SBO) and anhydrous milk fat (AMF). Under the investigated conditions (storage temperature = 20°C), the SBO remained in a liquid state, whereas fraction of the AMF crystallized. As also shown by other researchers, it was found that the longer-chain length SE surfactant promoted the AMF crystallization leading to the formation of smaller in size and larger in number crystallites, whereas the shorter lauric SE inhibited the crystallization of AMF and promoted the formation of larger AMF spherulites. Accordingly, the emulsions prepared with P170 exhibited increased stability compared to the emulsions prepared with L195. Furthermore, the authors found that at least 30 wt. % AMF were needed to be present in the SBO+AMF oily mixture to formulate w/o emulsions which remain stable against gravitational separation upon storage. This result was attributed to the ability of AMF to crystallize and form network which supports the water droplets. The results from the experiments with varying the concentration of L195 SE, showed that the emulsions prepared with lower surfactant concentrations were more stable compared to those prepared with higher L195 concentration. This was related to the difference in the AMF crystalline network properties, which was bulkier when higher surfactant concentrations were used, and thus unable to fix the water droplets [66]. The emulsification process in this study was performed in the one-phase region, i.e. at high temperature where the SE is melted.

Another approach was used in Ref. [67*] to prepare w/o emulsions, stabilized by sucrose stearate (C1801, HLB = 1). In this study, instead of preparing the initial emulsions at high temperature and then cooling them down to temperatures below the SE phase transition temperature, the w/o emulsions were directly prepared at low temperature. To do this, the authors initially prepared the isotropic SE-in-oil solutions at high temperature, then cooled them to low temperature, allowing crystallization of the surfactant molecules in 1-8 µm spherulitic crystals. Afterwards, water was added to this SE-in-oil dispersion and emulsified in it at ambient temperature. The SE crystals served as particles which adsorbed on the water droplets surfaces, thus providing a Pickering stabilization against coalescence. Emulsions with high water weight fractions, $\varphi_w = 0.5$, were successfully prepared using this approach, providing that the surfactant concentration was at least 2 wt. %. At $\varphi_w = 0.75$, the prepared high internal phase emulsions had gel-like properties, thus they remained stable even when lower surfactant concentrations were used. The emulsions had excellent long-term stability (up to 6 months) irrespectively of the oil used (hexadecane, canola oil, olive oil or SBO) when stored at temperatures which were higher than the oil crystallization temperature, but lower compared to the SE melting temperature. However, when complete or partial crystallization of the continuous phase was allowed, or alternatively – the temperature was raised to allow melting of the interface crystals, the emulsions became partially or fully destabilized, making them temperature responsive [67*].

Combination of SE with beeswax was also found to be appropriate for stabilization of w/o emulsions with relatively high dispersed phase fraction, $\varphi_w = 0.6$ [68]. The best results were obtained when 1.5 wt. % sucrose ester (HLB = 1) was combined with 0.5 wt. % beeswax, which ensured formation of both spherical and fine crystals, which stabilized the water droplets against aggregation and coalescence. The increase of the concentration of the high melting temperature beeswax was found to increase the elastic modulus of the prepared viscoelastic emulsions, but in an expense for a shrink linear viscoelastic region, i.e. these emulsions were more susceptible to damage by external deformations [68]. Solely non-covalent interactions were suggested to exist between the SE and beeswax molecules, as the studied infrared spectra did not show appearance of new characteristic peaks when the two surfactants were dissolved in the oil [68].

In contrast, formation of complex via H-bonds between the sucrose stearate (S170) and soybean phosphatidylethanolamine (SPEA) was demonstrated in Ref. [69**]. The complex was prepared by dissolving the SE and SPEA surfactants in chloroform and then evaporating it. Particles with finely tuned properties, which were combination of the properties of S170 and SPEA, were obtained depending on the ratio between the two surfactants. In particular, this was used to alter the hydrophobicity of the S170 particles, which were very hydrophobic showing a three-phase contact angle of 136°, while the SPEA particles were hydrophilic with three-phase contact angle of 66.5°. The prepared complexes had intermediate hydrophobicities, but the authors found that the complex prepared with 3:1 ratio between SE:SPEA was the most appropriate one for stabilization of w/o HIPEs. In this case, the measured three-phase contact angle was ≈ 118°. Furthermore, this complex was found to be able to decrease the water-oil interfacial tension the most (down to ca. 1.5 mN/m), while ensuring the highest viscoelasticity for the interfacial adsorption layer, G' ≈ 25 mPa, whereas for the SE or SPEA alone the interfacial G' ≈ 4-5 mPa [69**]. The prepared HIPEs ($\varphi_w$ = 0.75) were gel-like and remained macroscopically stable for 1 month, although some drop size increase was observed. We note that similar alteration in the three-phase contact angles for SE was also observed when their hydrophobic tail was varied in Ref. [30], see also the last paragraph in Section *Sucrose esters in solid state* above. The best w/o HIPEs prepared in this study were obtained with sucrose oleate, which had a three-phase contact angle ≈ 101° [30].

## 6. Foamed emulsions (foamulsions)

The crystallization ability of the SEs upon cooling is also widely used for preparation of stable foamed emulsions (foamulsions) [18,70-74]. Foamulsions are broadly spread in food industry, e.g. whipped creams and ice-creams are foamulsions [74,75]. Typically, these products are prepared in two stage process – first the oil-in-water emulsion is prepared. Then, it is usually aged for a given period of time at fridge temperature. The final product is obtained after an intensive stirring step in which air bubbles become incorporated inside the prepared emulsions. Typically, the foamulsions are stabilized by combination of protein and low-molecular weight emulsifiers, in addition to the frozen network comprised of partially coalesced oily droplets.

The effect of addition of SEs with various hydrophobicity on the crystallization of bulk oils (hydrogenated palm kernel oil, HPKO and anhydrous milk fat, AMF), their ability for preparation of oil-in-water emulsions and their whipping capabilities were studied in Refs [70-73]. In all cases, the authors found that the SEs with saturated FA residues had greater ability to induce surface and bulk nucleation at higher temperatures compared to SEs with shorter or unsaturated tails, similarly to the conventional low molecular weight surfactants [76,77]. The short- or unsaturated-chain SEs did not affect the initial crystallization temperature of the studied triglyceride oil, but impeded the crystalline growth, because they remained in liquid state, leading to lower temperatures for oil crystallization [70,71]. The more hydrophilic SEs with the same long saturated hydrophobic tail (usually $C_{16}$-$C_{18}$), i.e. with higher monoester content, were found to be able to displace the protein molecules from the interface to the largest extent, thus leading to significantly decreased interfacial tension [72,73].

For example, the results obtained in Ref. [70] demonstrated that the sucrose ester S170 was able to induce faster nucleation in the fat blend AMF:HPKO = 3:2, which ensured a good stabilization of the prepared whipped cream. In contrast, when the unsaturated SE O170 was used (with oleic tail), it increased the emulsion drop sizes, decelerated the crystallization rate, thereby decreasing the firmness of the obtained crystalline network and stability of the whipped creams.

The effects of alkyl chain length for SEs with HLB = 1 and ME concentration for stearoyl SEs were studied in Ref. [71] for HPKO whipped creams. It was shown that the presence of L195 and O170 in HPKO leads to formation of larger crystals, while the stearoyl SEs with ME content between 5 % and 25 % (S170, S370 and S570) served as nucleation agents and led to formation of fine and homogeneously distributed crystals inside the oil. As a consequence, the fat network formed with L195 and O170 was weaker and the prepared whipped creams exhibited lower stability, whereas the stearoyl SEs formed strong crystalline network which was able to stabilize the incorporated air bubbles inside it. The best results are obtained when S170 is used [71].

The effect of ME content in stearoyl SEs was studied in Ref. [72] for whipped creams prepared from AMF only. It was demonstrated that foamulsions formulated with AMF in presence of more than 0.125 wt. % S370 SE were completely unstable, whereas much higher stability was demonstrated for foamulsions prepared with the more hydrophilic S770 and S1670 SEs in the concentration region between 0.125 and 0.75 wt. %. The instability of the foams, prepared with higher S370 concentrations, was attributed to the increase of drop sizes in the initial emulsions, average sizes around 3.4 μm for 0.125 wt. % and 8.9 μm for 0.75 wt. % were reported [72].

The ability of SE to increase the aqueous solutions viscosity by forming a network of supramolecular structures, see Section *Sucrose esters self-assembly in aqueous media* above, was used by Zheng et al. to prepare stable water-continuous foamulsions with liquid vegetable oils (canola oil and soybean oil) in presence of sucrose stearate C1807 (HLB ≈ 7, 40% monoester content) [18]. Samples prepared with 3 wt. % and higher surfactant concentrations in presence of 50 wt. % oil appeared as gels and remained completely stable for more than 2 months. However, the overrun remained unchanged (ca. 35%) only when the SE concentration was 4 wt. %, whereas it was significantly decreased in the 3 wt. % sample. The increase of oil content was found to be also unfavorable regarding the foamulsions stability [18].

7. Oleogels

Several studies have explored the opportunity to use SEs as oleogelators, which should be able to structure liquid vegetable oil into a three-dimensional network forming an oleogel [78-81]. Oleogels have attracted significant interest in recent years due to their potential for usage as alternative to saturated and trans fats usage in the lipid-based food products, such as margarine, ice cream, chocolate, fillings, etc. [81]. The investigations of the SEs using the traditional oleogel formation route, i.e. via homogenization of the SEs and oil at high temperature, followed by the cooling of the obtained solutions, concluded that relatively high concentrations of lipophilic SEs are needed (20 wt. % S270 or S370) to form a stable oleogel with MCT oil [78].

In a later study, da Silva et al. investigated the possibility for oleogels preparation with only 10 wt. % SEs in rapeseed oil [80*]. Three different preparation routes were tested for introduction of the SE into the oil – the *traditional* one, including direct mixing at high temperature; *ethanol route*, which comprised pre-dissolving the studied SE into ethanol and then mixing it with pre-heated oil at high temperature and evaporation of the ethanol; and the so-called '*foam-template method*' which included dispersion of the studied SEs into water, then its homogenization and freeze-drying for 3 days at -80°C into powder which was then introduced into the oil at room temperature in order to produce an oleogel. Stearic-palmitic sucrose esters with variable content of the monoesters (between 10 and 70 wt. %) were investigated.

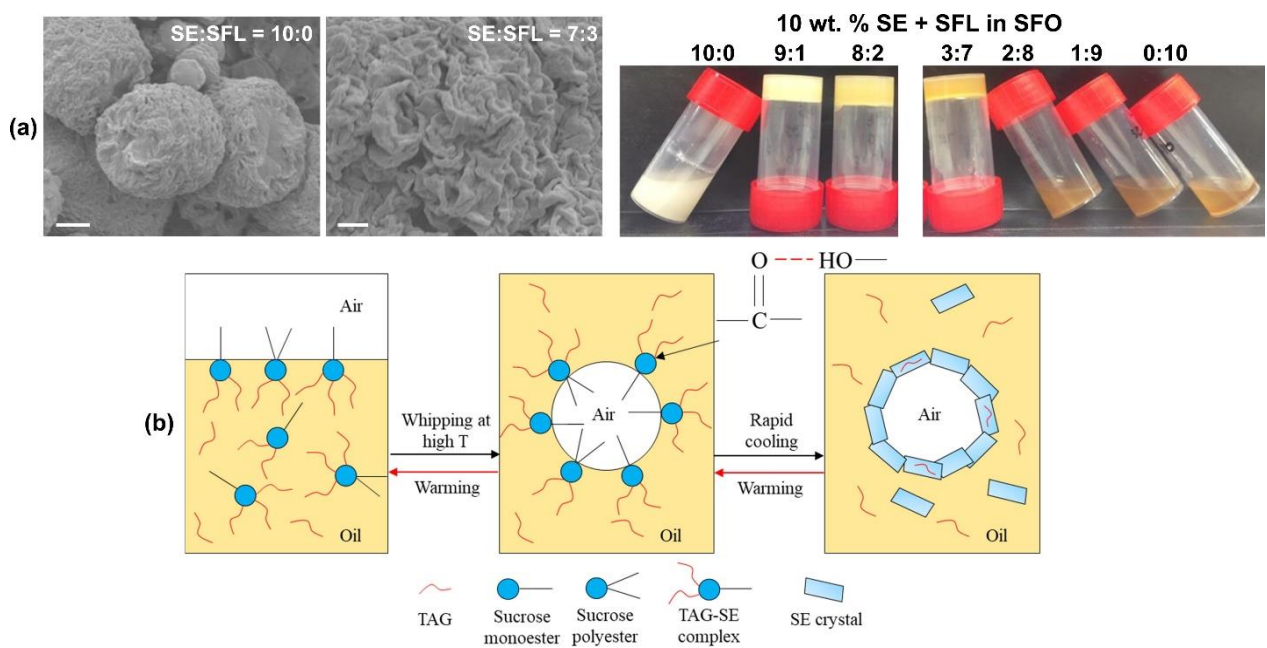

**Figure 5.** Effect of formation of H-bonds between sucrose esters and other type of molecules. **(a)** Oleogels formed from 10 wt. % sucrose ester (SE) + sunflower lecithin (SFL) dissolved in sunflower oil in various ratios. Gels are obtained for SE:SFL = 9:1 to 3:7 ratios, whereas the other samples are viscous liquids at 25°C. The cryo-scanning electron microscopy images illustrate the oleogels structure for SE:SFL = 10:0 and 7:3 ratio. Scale bar is 1 μm. Adapted and reprinted with permission from Ref. [79], copyright 2017 © Elsevier. **(b)** Mechanism proposed to explain the formation of stable oleofoams at 80°C with stearic SE (HLB ≈ 5) and triglyceride oils. Formation of SE-triglyceride complex is suggested via H-bonding, which allows bubble stabilization via adsorption of the SEs at the oil-air interface. Reproduced with permission from Ref [82**], copyright 2021 © Elsevier.

The most promising results were obtained for the SE with intermediate hydrophilicity (SP50, stearate/palmitate sucrose esters with HLB ≈ 11 and 50% monoesters content) when the *ethanol route* was used [80*]. With this sample, an oleogel with hardness of ca. 0.35 N and elastic modulus $G'$ ≈ 4.6 kPa was successfully prepared. It contained two interconnected types of crystal morphologies – needles and globular crystals. The authors did not find significant differences in the physicochemical properties of this and the other studied samples, including their melting temperatures, molecular arrangement studied by SAXS/WAXS, and infrared spectra, although stable oleogel with good mechanical properties was observed only in this case. Thus, the obtained result was attributed to the ability of ethanol to enhance the dissolution of the SE into the oil by improving the dissolution of the monoesters, which then formed higher number of hydrogen bonds with oily molecules and entrapped them into a stable oleogel. Sucrose ester SP70 (sucrose stearate/palmitate with HLB ≈ 15 and 70% monoesters) was also investigated. However, its more hydrophilic nature interrupted its dissolution into the oil via the *traditional* oleogel formation method, whereas when the *ethanol* method was used, only globular crystals were formed, which were unable to entrap the whole rapeseed oil. The best oil-binding capacity was demonstrated for the oleogel with SP70 prepared by the *foam-template* method, but its hardness remained quite low (< 0.05 N) [80*].

Other studies demonstrated improved oleogel formation ability for SEs mixed with additional lipophilic emulsifier, e.g. lecithin or monoglycerides [79,80*]. For example, SFO oleogel with storage modulus $G' \approx 20$ kPa was prepared using 10 wt. % oleogelator comprising hydrophobic sucrose ester (stearic acid based, HLB ≈ 2 with ca. 10 % monoester content) and sunflower lecithin (SFL), mixed in 7:3 ratio [79], see Figure 5a. The sucrose ester and the lecithin *per se* did not allow formation of stable oleogel, Figure 5a. Therefore, the synergistic effect observed when both emulsifiers were added was attributed to the hindered formation of hydrogen bonds between the SEs monomers, which interrupted their enclosure into globular structures. Instead, when lecithin was present, it interacted with the SEs through hydrogen bonding and helped the formation of dense uniform crystalline network, which spread throughout the whole sample, thus entrapping the liquid triglyceride molecules inside it [79]. Combination between stearic/palmitic hydrophilic sucrose ester (HLB ≈ 15, 70% monoesters content) with saturated monoglycerides or fully hydrogenated rapeseed oil at 10 wt. % level (1:1 mixture), was also found to be suitable for preparation of stable oleogels with good mechanical properties using the *foam*-templated approach [81].

These results demonstrate the potential of SEs to be used as structuring agents for oils, which has proven to be possible, especially when they are combined with a second emulsifier. However, while some researchers demonstrated better results with SEs with lower monoester content, in other – the SEs containing higher monoester fraction were found to be more appropriate for formation of oleogels. Currently, the general mechanism behind these findings remain unknown. Therefore, further efforts are needed to provide guiding rules about an easy selection of the most appropriate oleogel formation route, the SEs which should be used and the exact mechanisms, determining the mechanical properties of the prepared oleogels.

## 8. Oleofoams

Another intriguing usage of SEs and their temperature-responsive properties was demonstrated in the study of Liu and Binks [82**]. These authors explored the possibility for formation of stable foams in which the continuous phase was comprised of oil molecules (oleofoams).

A stearic SE surfactants with HLB ≈ 5 and 30% monoester content was employed in this study. The surfactant was dissolved either in extra virgin olive oil, high oleic sunflower oil, rapeseed oil or refined peanut oil, which was then subjected to a whipping, using a hand hold electrical whisk under various conditions. The authors were able to prepare stable foams with overrun varying between ca. 160% for the olive oil and refined peanut oil at 5 wt. % SE content and up to 330 wt. % for refined peanut oil at 10 wt. % SE content. This successful foaming was only possible at temperature higher than the one at which the SE surfactant starts to crystallize or to form aggregates inside the oil, i.e. at $T \approx 80°C$. A little or no foaming was observed at lower temperatures [82**]. Note that this result is highly unexpected, as usually the other studies of oil foams have previously demonstrated much better results for systems in which lipid crystals are already present in the formulation in contrast to systems comprising a one-phase molecular solutions [82**-86].

The foams stability was also evaluated in Ref. [82**]. An almost complete destruction of the generated foams was observed in about 5 days when they were stored at 80°C. In contrast, foams with excellent stability and unchanged bubble sizes even after a few months' storage were demonstrated when the hot oil foams, prepared at 80°C, were quickly cooled to -5°C immediately after the foaming ended, and then were stored in a fridge at $T \approx 7°C$. The bubbles in the stable foams were with non-spherical shapes, demonstrating that they contained a crystalline shell of surfactant, providing a Pickering type of stabilization. Furthermore, these foams were found to be thermo-responsive, as they begin to destroy upon heating.

A mechanism, explaining these non-trivial observations was proposed by Liu and Binks [82**]. In particular, the good foamability in the one-phase region, where the SE was completely dissolved in the oil as compared to the poor foaming of the same system at lower temperatures was explained with three main factors: (1) the SEs monomer concentration decreases when the temperature is decreased as the SE molecules arrange either into micelles or into some other supramolecular crystalline structures; (2) the spontaneous adsorption of the precipitated SE crystals at the oil-air interface which is needed to stabilize the entrapped air bubbles was estimated to be energetically unfavorable, because their solid-air surface energy is considerably higher compared to the energy of the bare oil-air interface ($\sigma \approx 43$ mN/m *vs* 33 mN/m); (3) the crystals formed at lower temperatures may potentially have antifoaming effect as previously demonstrated in other studies [87]. Furthermore, the surface tension measurements performed at 80°C with the olive oil and SE mixture demonstrated significantly lower values, $\sigma \approx 23$ mN/m, compared to the interfacial tension measured at lower temperatures, $\sigma \approx 33$ mN/m. Note that the surface tension of neat olive oil was $\sigma \approx 26.5$ mN/m at 80°C. Therefore, a formation of surface active complex via hydrogen bonds between the carbonyl groups in the triglyceride molecules and the hydroxyl groups in the SEs was suggested which is responsible for the facilitated foam formation in the one-phase region at high temperatures, see Figure 5b [82**]. However, as the adsorbed molecules can easily exchange with those in the bulk oily phase, the prepared foams were unstable at 80°C, whereas the freezing of SE molecules upon cooling formed solid shell around the bubble surfaces, thus preventing their coalescence and preserving the foam stability.

## 9. Conclusions and outlook

The present review highlights the broad versatility of sucrose esters across diverse colloidal systems. Remarkably, these surfactants exhibit behavior that is comparable to, or even superior to the one known for conventional nonionic alcohol ethoxylates. Additionally, a significant advantage of SEs is their biodegradability and ease of production from sustainable sources.

Three key factors must be considered when selecting the most suitable SE for a specific application:

(1) The composition of mono-, di- and polyesters within the SE, which determines the hydrophilic-lipophilic properties of the surfactant. Typically, SEs with higher monoester content are used for water-continuous systems, while an increase in the amount of di- and higher esters is beneficial for oil-continuous formulations. However, the presence of a moderate concentration of diesters (up to ca. 50%) alongside monoesters often enhances the long-term stability of aqueous foams and emulsions. Additionally, intriguing rheological properties have been observed due to the formation of mixed wormlike micelles between the sucrose mono- and diesters.

(2) The chain length of SEs determines their phase behavior, which significantly impacts the generation of foams and emulsions, as well as their stability during storage. Specifically, the use of SEs with fluidized tails is usually advantageous for the formation of aqueous foams and emulsions. Conversely, the solidification of surfactant tails following the completion of the emulsification/foaming process typically enhances the long-term stability of these systems. In contrast, for oil-continuous formulations, surfactants with frozen tails are usually preferred. This is because they may induce oil nucleation and growth into smaller crystals or provide crystals which may adsorb on the newly created interface, thereby offering Pickering-type stabilization.

(3) Sucrose esters possess multiple hydroxyl groups capable of forming of H-bonds. Typically, these bonds form between individual SE molecules, providing a highly ordered molecular arrangement even at very high temperatures in bulk SEs or promoting easier molecular alignment in solutions. However, several studies have demonstrated that SEs may also engage in H-bonds with triglyceride molecules or co-surfactants, thus altering the hydrophobicity of SEs. This ability can be advantageous for formulations which cannot be generated otherwise (e.g. stable oleogels, oleofoams or water-in-oil emulsions).

While significant progress is already made in elucidating the details in the intricate behavior of SEs, numerous unresolved questions remain, necessitating further investigation. For instance, the precise structure of SE entities formed in solutions and how this structure depends on the degree of SE esterification, concentration, temperature, and the presence of additives (such as carbohydrates and other surfactants). Another interesting question is what is the origin of the negative zeta potential in SE systems reported in the literature, along with the potential for controlling and utilizing it for specific applications. Additionally, although the SEs have been shown to alter the triglyceride crystallization, the molecular mechanism underlying this phenomenon remains unknown, as does the rationalization of this effect for triglyceride molecules with varying structures. A comprehensive understanding of the mechanisms driving these and other scientific questions outlined in the paper will significantly help in establishing comprehensive guidelines for selecting the most appropriate sucrose ester surfactant for specific applications, making them excellent candidates for sustainable, functional ingredients.

**Funding:** This study is financed by the European Union-Next Generation EU, through the National Recovery and Resilience Plan of the Republic of Bulgaria, project № BG-RRP-2.004-0008-C01.

**Diana Cholakova:** Investigation, Visualization, Formal Analysis, Writing: Original manuscript, Review & Editing, Conceptualization; **Slavka Tcholakova:** Visualization, Writing: Review & Editing, Conceptualization, Supervision.

*Comprehensive overview of the main factors affecting the polymorphic phase transitions in triglycerides and their mixtures.*

*This work demonstrates the formation of ultra-stable foams (for more than a year) from a mixture of sucrose monoester and diester dissolved in 75 wt. % glucose syrup. This was explained with the formation of an insoluble, self-assembled surfactant layer which covered the surface of the microbubbles and induced elastic response of the interface, arresting the shrinkage of the bubbles.*